\documentclass[namedreferences]{SolarPhysics}
\pdfoutput=1
\usepackage[optionalrh]{spr-sola-addons} 
\usepackage{graphicx}        
\usepackage{amssymb}         
\usepackage[usenames,dvipsnames]{color}          
\usepackage{url}             

\DeclareFontFamily{OT1}{pzc}{}
\DeclareFontShape{OT1}{pzc}{m}{it}%
             {<-> s * [1.1500] pzcmi7t}{}
\DeclareMathAlphabet{\mathscr}{OT1}{pzc}%
                                 {m}{it}

\usepackage{epstopdf}
\usepackage{amssymb,amsmath}

\newcommand{\half}{{\textstyle\frac{1}{2}}}

\newcommand{\re}{\mathop{\rm Re}\nolimits}
\newcommand{\im}{\mathop{\rm Im}\nolimits}

\newcommand{\Ai}{\mathop{\rm Ai}\nolimits}
\newcommand{\Bi}{\mathop{\rm Bi}\nolimits}

\newcommand{\rd}{\mathrm{d}}

\newcommand{\pderiv}[2]{\frac{\partial#1}{\partial#2}}
\newcommand{\pderivd}[2]{\frac{\partial^2#1}{\partial#2^2}}

\newcommand{\deriv}[2]{\frac{\rd#1}{\rd#2}}

\renewcommand{\k}{\mathbf{k}}
\newcommand{\x}{\mathbf{x}}
\newcommand{\B}{{\mathbf{B}}}

\newcommand{\F}{{\mathbf{F}}}

\newcommand{\U}{\mathbf{U}}
\newcommand{\V}{\mathbf{V}}

\newcommand{\D}{\mbox{\boldmath$\mathsf{D}$}}
\newcommand{\R}{\mbox{\boldmath$\mathsf{R}$}}
\renewcommand{\S}{\mbox{\boldmath$\mathsf{S}$}}
\newcommand{\Q}{\mbox{\boldmath$\mathsf{Q}$}}

\newcommand{\vdot}{{\boldsymbol{\cdot}}}

\newcommand{\grad}{\mbox{\boldmath$\nabla$}}
\newcommand{\bxi}{\mbox{\boldmath$\xi$}}

\newcommand{\diag}{\mathop{\rm diag}}
\newcommand{\thth}{\hspace{1.5pt}}

\newcommand\Div{\grad\vdot\thth}

\newcommand{\kperp}{k_{\scriptscriptstyle\!\perp}}
\newcommand{\kperpz}{k_{\scriptscriptstyle\!\perp0}}

\newcommand{\ri}{\mathrm{i}}

\newcommand{\calL}{{\mathcal{L}}}
\newcommand{\calD}{{\mathcal{D}}}

  \renewcommand{\le}{\leqslant}

\newcommand{\aap}{\textit{Astron. Astrophys.}}

\newcommand{\apj}{\textit{Astrophys. J.}}

\newcommand{\jgr}{\textit{J. Geophys. Res.}}
\newcommand{\mnras}{\textit{Mon. Not. Roy. Astron. Soc.}}

\newcommand{\solphys}{\textit{Solar Phys.}}

\allowdisplaybreaks[1]

\begin{document}
\begin{opening}

\title{Resonant Absorption as Mode Conversion?}

\author{P.~S.~\surname{Cally}$^1$\sep
J.~\surname{Andries}$^{1,2}$}

\institute{$^1$ Centre for Stellar and Planetary Astrophysics, School of
Mathematical Sciences, Monash University, Victoria 3800, Australia \email{paul.cally@monash.edu} \email{jesse.andries@monash.edu}\\
$^2$ Centrum voor Plasma-astrofysica, K.U.Leuven, Celestijnenlaan 200B,
B-3001 Leuven, Belgium
\email{jesse.andries@wis.kuleuven.be}
}

\runningauthor{P.S. Cally and J. Andries}

\runningtitle{Resonant Absorption as Mode Conversion?}

\begin{abstract}
Resonant absorption and mode conversion are both extensively studied mechanisms for wave ``absorption'' in solar magneto\-hydro\-dynamics (MHD). But are they really distinct? We re-examine a well-known simple resonant absorption model in a cold MHD plasma that places the resonance inside an evanescent region. The normal mode solutions display the standard singular resonant features. However, these same normal modes may be used to construct a ray bundle which very clearly undergoes mode conversion to an Alfv\'en wave with no singularities. We therefore conclude that resonant absorption and mode conversion are in fact the same thing, at least for this model problem. The prime distinguishing characteristic that determines which of the two descriptions is most natural in a given circumstance is whether the converted wave can provide a net escape of energy from the conversion/absorption region of physical space. If it cannot, it is forced to run away in wavenumber space instead, thereby generating the arbitrarily small scales \emph{in situ} that we recognize as fundamental to resonant absorption and phase mixing. On the other hand, if the converted wave takes net energy way, singularities do not develop, though phase mixing may still develop with distance as the wave recedes.
\end{abstract}



\end{opening}


\section{Introduction}
The conversion of MHD waves from one type to another has attracted much attention over many years. Recently, \inlinecite{sc06} have adapted the dispersion-relation-based generalized ray theory formalism of \inlinecite{tkb03} to the problem of conversion between fast and slow waves, finding that the ``fast-to-slow transmission'' occurring near the Alfv\'en/acoustic equipartition level depends crucially on the attack angle made by the incident wave vector to the magnetic field. Comparison with exact solutions in an isothermal atmosphere shows that this performs very well indeed at small to moderate attack angles \cite{hc09}. Fast-to-Alfv\'en conversion has not yet been satisfactorily described in this manner, though a wave-mechanical approach shows that it can be very significant in certain attack directions \cite{cg08}.

The Generalized Ray Theory (GRT) of \inlinecite{tkb03} and \inlinecite{sc06} reduces the problem of mode conversion/transmission to a local analysis of a saddle point in an appropriate phase space. This can be done in a quite general way, allowing the development of a simple recipe for calculating the transmission and conversion coefficients.

Parallel and seemingly disconnected theoretical developments in MHD wave theory have focussed on the well known phenomenon of resonant absorption, in which a wave is ``absorbed'' on a certain resonant surface (see \opencite{gp04} for extensive discussion). Applications to solar physics are numerous, ranging from theories of p-mode absorption by sunspots \cite{holl88}, to mechanisms for thermalization of Alfv\'en waves in coronal loops \cite{ion78} and the damping of TRACE loop oscillations \cite{rr02}. The physical mechanism by which resonant absorption operates through phase mixing to absorb surface waves is brought out clearly by \inlinecite{cs92}. 

But are mode conversion and resonant absorption always distinct processes? In this paper we look at perhaps the simplest resonant absorption model in which a fast wave approaches and is absorbed by an Alfv\'en resonance in a cold MHD plasma (plasma-$\beta=0$). Now, an ``Alfv\'en resonance'' can also be interpreted as just an Alfv\'en wave running along a particular field line (or flux surface), and so can also be thought of as a result of mode conversion. Our aim here is therefore to better understand the relationship between the two theories. In this way, we hope to build a bridge between mode conversion and resonant absorption theory, or at least to establish a bridgehead.

\section{Modal Model and Governing Equations}    \label{sec:model}
Consider a cold (sound speed zero) magnetoatmosphere with uniform magnetic field $\B$ in the $z$-direction (cartesian coordinates $(x,y,z)$), but where density $\rho$ varies with $x$ only. Assume an ideal magneto\-hydro\-dynamic linear wave with plasma displacement vector of the form $\bxi(x,y,z,t)=(\xi_x(x),\xi_y(x),0)\exp[\ri(k_y y + k_z z-\omega t)]$; the only restoring force is perpendicular to $\B$, so there can be no $\xi_z$. The governing differential equations are then
\begin{equation}
\begin{aligned}
\left(\pderivd{}{x}+\frac{\omega^2}{a^2}-k_z^2\right)\xi_x &= -\ri k_y \pderiv{\xi_y}{x}, \\[4pt]
\left(\frac{\omega^2}{a^2}-k_y^2-k_z^2\right)\xi_y &= -\ri k_y \pderiv{\xi_x}{x},
\end{aligned}
\ \ 
\label{xiDEs}
\end{equation}
where $a(x)$ is the Alfv\'en speed, showing that $\xi_x$ and $\xi_y$ decouple to independent fast ($\xi_x$) and Alfv\'en ($\xi_y$) waves if $k_y=0$. The associated WKB dispersion relation is
\begin{equation}
\calD = \left(\frac{\omega^2}{a^2}-k_x^2-k_y^2-k_z^2\right)\left(\frac{\omega^2}{a^2}-k_z^2\right)=0\, ,     \label{D}
\end{equation}
where $k_x$ is the WKB wavenumber associated with $x$. The two factors represent the fast and Alfv\'en modes respectively.

Equations (\ref{xiDEs}) may be easily combined into a single second order ordinary differential equation for the dilatation $\chi=\Div\bxi=\partial\xi_x/\partial x+\ri\, k_y\xi_y$ (see also Equation (1) of \opencite{holl90}):
\begin{equation}
\epsilon^2\deriv{\ }{x}\left(\frac{1}{\beta}\deriv{\chi}{x}\right) + \frac{\beta\cos^2\theta-\sin^2\theta}{\beta}\,\chi=0\, ,         \label{DE}
\end{equation}
where $\beta=\beta(x)=\omega^2/(a^2k_z^2)-1=\Omega^2\sec^2\theta-1>-1$, $k_y=\kperp\sin\theta$, $k_z=\kperp\cos\theta$, and $\Omega=\omega/a\kperp$. Here, $\kperp=\sqrt{k_y^2+k_z^2}=\epsilon^{-1}$ is the component of the wavevector perpendicular to the direction of stratification $x$. The corresponding components of displacement are $\xi_x=-k_z^{-2}\beta^{-1}\chi'=-\epsilon^2\beta^{-1}\sec^2\theta\, \chi'$ and $\xi_y=-\ri\,\beta^{-1}k_y k_z^{-2}\chi=-\ri \,\epsilon\,\beta^{-1}\sin\theta\sec^2\theta\,\chi$.
The associated wave energy (Poynting) flux is
\begin{equation}
\mathbf{F} = F_0\left(
\frac{\epsilon^2\sec^2\theta}{\beta} \im\left(\chi'\chi^*\right),
\frac{\epsilon}{\beta} \sin\theta\sec^2\theta\, |\chi|^2,
\frac{\cos\theta}{\epsilon}|\bxi|^2
\right)
\, ,   \label{F}
\end{equation}
where $F_0=\omega B^2/\mu =\omega\,\rho\, a^2$, $B=|\B|$ is the background magnetic field strength and $\mu$ is the magnetic permeability. Note that $F_0$ has the dimensions of a flux per unit length, or in other words an energy density per unit time. By energy conservation (or by reference to the Wronskian of the solutions of Equation (\ref{DE})), the $x$-component $F_x$ is constant, except for a discontinuity at $\beta=0$ which is the signature of resonant absorption. The corresponding wave energy density is
\begin{equation}
E = \half\rho\left[ (\omega^2+a^2k_z^2)|\bxi|^2+a^2|\chi|^2\right] \,.  \label{E}
\end{equation}

We shall assume that $\beta$ \emph{decreases} monotonically with $x$. Assume a rightward propagating fast magneto\-acoustic wave classically reflects at $x=0$, \emph{i.e.}, $\beta(0)=\beta_0=k_y^2/k_z^2=\tan^2\theta$, and that there is an Alfv\'en resonance at $x=X>0$, \emph{i.e.}, $\beta(X)=0$. An incident fast wave in Region I ($x<0$) reflects in Region II ($x\approx0$), but partially tunnels through Region III ($0<x<X$) to be resonantly absorbed in Region IV ($x\approx X$). There remains only an evanescent tail in Region V ($x>X$). If $k_y=0$ ($\theta=0$), Region III collapses to zero width and Regions II--IV coalesce. 

With these points in mind, and using the series expansion (\ref{froblead}) from Appendix \ref{app:gen}, the total $x$-component of wave energy flux may be shown to be
\begin{equation}
F_x = \pi\,h\,F_0\tan^2\theta\,\left| \chi(X)\right|^2\,
\mathcal{U}(X-x)\, ,  \label{Fx}
\end{equation}
where $h=-1/\beta'(X)>0$ and $\mathcal{U}$ is the unit step function. $\chi$ is everywhere continuous so there is no need to distinguish between $X^+$ and $X^-$ here.

Equation (\ref{E}) may be used to define a net energy transport velocity $\V=\F/E$. Employing the Frobenius series again it is easily confirmed that $\V = (\mathcal{O}((x-X)^2)\,\mathcal{U}(X-x), \mathcal{O}(x-X),\, a + \mathcal{O}(x-X))$ near the resonance $X$ (where $|\bxi|\to\infty$ and $\chi$ is finite). Note that it takes an infinite time to reach the resonance at speed $V_x$.

\section{Mode Conversion}   \label{conv}

Mode conversion is the process whereby one clearly identifiable wave type changes into another type in some (normally) compact conversion region where the distinction between the modes breaks down. However, away from this region, the incident, converted and transmitted waves are easily identified, and conversion/transmission coefficients assigned. The precise mathematical description of this process and approximate calculation of conversion coefficients may make certain assumptions to render the problem tractable, though the wider concept of mode conversion does not depend on these.

One such description is the so-called modular theory developed over the last two decades by many authors, summarized by \inlinecite{tkb03}, and adapted to MHD by \inlinecite{cally06}. It fundamentally assumes the wavelengths involved are large enough to fit several into the interaction region, a point which limits its applicability in the current context, and to which we shall return later in this section.

Modular theory is based on an action principle with action $\mathcal{A}=\int \calL\,d\x\,dt$ and Lagrangian density $\calL=\bxi^H\D\bxi$, where the superscript ``$H$'' denotes the Hermitian transpose. For the problem at hand here, the dispersion tensor is simply
\begin{equation}
\D = 
\begin{pmatrix}
\displaystyle
\frac{\omega^2}{a^2}-(k_x^2+k_z^2) & - k_x k_y \\[4pt]
- k_x k_y & \displaystyle\frac{\omega^2}{a^2}-(k_y^2+k_z^2)    \label{Dmat}
\end{pmatrix}.
\end{equation} 
Once again we assume $a^2=B^2/\mu\rho$ increases monotonically with $x$.

\begin{figure}[htbp]
\begin{center}
\includegraphics[width=.5\textwidth]{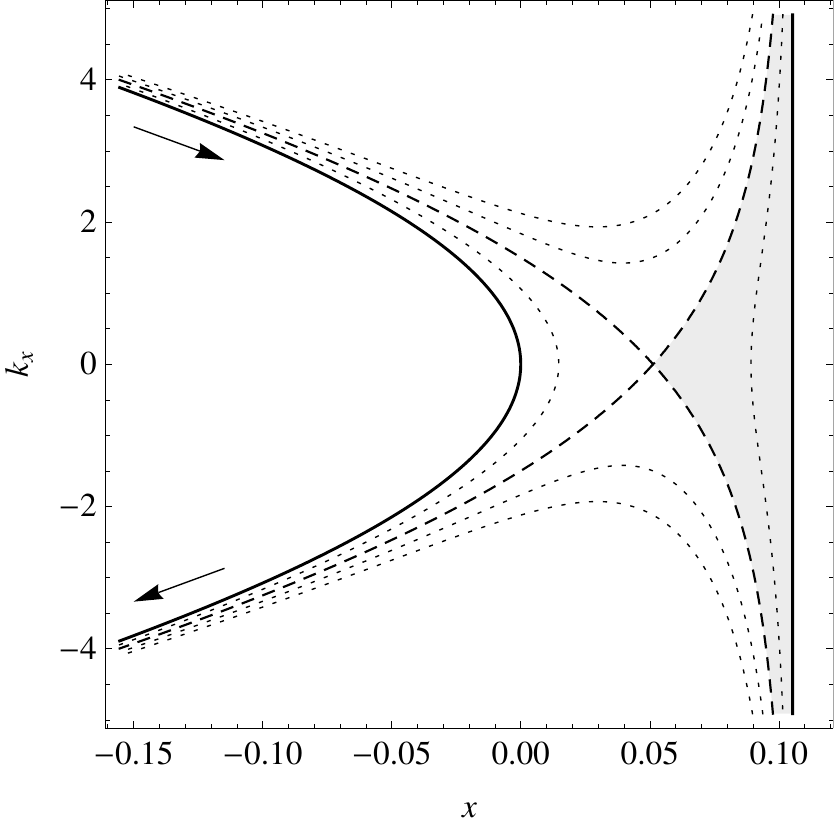}
\caption{Dispersion diagram for the case $k_y=3$, $k_z=9$ (see Section \ref{bundle}) with $\omega^2/a^2\kperp^2=e^{-x}$, for which $X=\ln(10/9)$ and $x_0=\ln(20/19)$. The heavy full curves correspond to $\calD=0$ and represent respectively the fast (left) and Alfv\'en (right) modes. The dashed curves are $\calD=\calD(x_\star,0)$, which pass through the saddle point. A few other values of $\calD$ are shown dotted. The arrows indicate the direction of energy propagation in the fast mode. The shaded region between the resonance and the principal saddle axes has area 0.31.}
\label{fig:ph}
\end{center}
\end{figure}

Fixing $k_y$ and $k_z$, contours of the dispersion function $\calD(x,k_x)=\det
\D$ may be plotted in the $(x,k_x)$ plane, displaying a saddle at the point $(x_\star,0)$, where $\omega^2/a^2=k_z^2+\half k_y^2$, \emph{i.e.}, $\beta_\star=\half\tan^2\theta-1$. As a concrete example (see Figure \ref{fig:ph}), we let $\Omega^2=\omega^2/a^2\kperp^2=e^{-x}$, for which $X=-\ln(k_z^2/\kperp^2)=\ln\sec^2\theta$, and the saddle point is located at $x_\star=-\ln((k_z^2+\half k_y^2)/\kperp^2)$, $k_x=0$.

Now, we may decompose $\D=\S\mathbf{\Lambda}\S^T$, where $\mathbf{\Lambda}=\diag[(\omega^2/a^2)-k_z^2,\,(\omega^2/a^2)-k^2]$ contains the two eigenvalues of $\D$, $k=|\k|$, and
\begin{equation}
\S = \frac{1}{k_p}
\begin{pmatrix}
-k_y & k_x \\
k_x & k_y
\end{pmatrix},
 \label{Smat}
\end{equation}
where $k_p=\sqrt{k_x^2+k_y^2}$ has been introduced for convenience.
The two columns of $\S$ are the normalized eigenvectors of $\D$. 

The mode conversion formalism of \inlinecite{tkb03} requires a different form though, in which the diagonal elements of the $2\times2$ dispersion matrix represent the asymptotic behaviour of the two mode types far from the interaction region. To that end, we introduce
\begin{equation}
\begin{split}
\tilde\D &=
\begin{pmatrix}
\frac{\omega^2}{a^2}-\frac{1}{2}k_p^2-k_z^2-\frac{1}{2}k_x\sqrt{k_x^2+2k_y^2} & \frac{1}{2}k_y^2 \\[4pt]
\frac{1}{2}k_y^2  & \frac{\omega^2}{a^2}-\frac{1}{2}k_p^2-k_z^2+\frac{1}{2}k_x\sqrt{k_x^2+2k_y^2}
\end{pmatrix}\\[8pt]
&=
\begin{pmatrix}
D_a & \tilde\eta \\
\tilde\eta^* & D_b
\end{pmatrix} = \R\mathbf{\Lambda}\R^T\,,
\end{split} \raisetag{16pt} \label{Dtilde}
\end{equation}
where
\begin{equation}
\R =  \frac{1}{\sqrt{2}\,k_p}
\begin{pmatrix}
\sqrt{k_p^2-k_x\sqrt{k_p^2+ k_y^2}} &
  \  -\sqrt{k_p^2+k_x\sqrt{k_p^2+ k_y^2}} \\[8pt]
\sqrt{k_p^2+k_x\sqrt{k_p^2+ k_y^2}} &
  \  \sqrt{k_p^2-k_x\sqrt{k_p^2+ k_y^2} }
   \end{pmatrix}\,.  \label{R}
\end{equation}

Both $\S$ and $\R$ are orthonormal matrices.
Here, the loci $D_a=0$ and $D_b=0$ correspond to the level curves of $\calD$ that pass through the saddle point, \emph{i.e.}, the dashed curves in Figure \ref{fig:ph}, with $D_a$ the incoming branch and $D_b$ outgoing. The off-diagonal terms $\tilde\eta=k_y^2/2$ are the coupling coefficients. As anticipated, there is no coupling when $k_y=0$. 

The point of all this is that we have now introduced a more suitable state vector $\U=\Q\bxi$ where $\Q=\S\R^T$ in place of $\bxi$. With it, the Lagrangian
\begin{equation}
\calL=\bxi^H \D \bxi = \U^H \tilde\D\U\, 
\end{equation}
remains unchanged in form though with the new dispersion tensor $\tilde\D$.
The components of $\U$ represent the amplitudes of the two decoupled (orthogonal) modes with eigenvalues $D_a$ and $D_b$ in the conversion region.
That a transformation leading to $\tilde\D$ is possible follows from Sylvester's Theorem (see \opencite{sc06} for details) and does not actually require explicit construction.

The classic ``transmission coefficient'' for energy ``transmitting'' from the fast to the Alfv\'en mode $T=\tau^2$ may then be calculated, where
\begin{equation}
\tau = \exp(-\pi |\eta|^2)_\star\, ,  \label{tau}
\end{equation}
with $\eta=\tilde\eta/|\mathcal{B}|^{1/2}$ and
\begin{equation}
\mathcal{B}=\{D_a,D_b\}=\pderiv{D_a}{k_x}\pderiv{D_b}{x}-\pderiv{D_a}{x}\pderiv{D_b}{k_x}=-2\kperp^2{\Omega^2}'\frac{k_x^2+k_y^2}{\sqrt{k_x^2+2k_y^2}}
\label{Poisson}
\end{equation}
is the Poisson bracket of the two uncoupled dispersion functions. The subscripted star indicates that $T$ should be evaluated at the star point $x=x_\star$, $k_x=0$. With the exponential profile $\Omega^2=e^{-x/h}$, and introducing the dimensionless parameter $\sigma=(\kperp h)^{2/3}\sin^2\theta$, it is then easily shown that
\begin{equation}
T = \exp\left[-\frac{\pi\sqrt{2}}{2}\, \frac{\sigma^{3/2}}{1+\cos^2\theta}\right]\, .  \label{T}
\end{equation}

Unfortunately, this result cannot be taken at face value. As stated earlier, modular theory relies on the solution being substantially wavelike (WKB) in the interaction region. The analysis of \inlinecite{cally05} makes clear how this operates in terms of stationary phase integrals; two neighbouring modes propagating in phase may efficiently exchange energy. A practical measure of whether modular theory applies is the (dimensionless) area $A_s$ in $x$-$k_x$ space of the region bounded by $D_a$, $D_b$, and the resonance (see the shaded region in Figure \ref{fig:ph}). It requires that $A_s\gg2\pi$ \cite{bkt08}, which fails spectacularly in the case of Figure \ref{fig:ph} ($A_s=0.31$). 
Nevertheless, comparison with numerical solutions in Section \ref{num} will reveal that $T$ always provides an upper bound on the wave-energy transmission to an Alfv\'en wave.

Lest the reader think that different or improved expressions for $T$ may be obtained by alternate choice of variables (\emph{e.g.}, arc length along a ray and the associated wavenumber), it should be born in mind that the independent variables $x$ and $k_x$ in Equation (\ref{Poisson}) must be canonical coordinates in the Hamiltonian sense, and that the Poisson bracket (and hence $T$) is invariant under arbitrary canonical transformations. The area $A_s$ is also invariant under canonical transformation, by Liouville's Theorem.

\section{Numerical Calculation of the Reflection and Absorption Coefficients} \label{num}
As befits a local analysis, we choose to approximate the arbitrary Alfv\'en profile by a suitable simple analytic form which is expected to suffice in general since mode conversion is indeed typically restricted to a compact region. The most convenient choice is to assume the density scale length $h$ is constant (its value at the resonance), whence $\Omega^2=e^{-x/h}$ and $\beta=e^{-(x-X)/h}-1$. Hollweg's \citeyear{holl90} choice of a linear profile is problematic, as it unphysically results in $\Omega^2<0$ for sufficiently large $x$ unless truncated by some \emph{ad hoc} and troublesome boundary. Unfortunately, this simple profile fails to produce an equation with known exact solution (as does the linear profile).

Introducing $\zeta=\kperp^{2/3}h^{-1/3}(x-X)$ as the new dimensionless independent variable, and the dimensionless parameters $\sigma=(\kperp h)^{2/3}\sin^2\theta$ and $\nu=(\kperp h)^{-2/3}=\sigma^{-1}\sin^2\theta$, Equation (\ref{DE}) takes the form
\begin{equation}
\deriv{\ }{\zeta}\left(\frac{1}{u(\zeta)}\deriv{\chi}{\zeta}\right) - \left(\cos^2\theta+\frac{\sigma}{u(\zeta)}\right)\,\chi=0       \label{ERAT}
\end{equation}
where $u=(1-e^{-\nu\zeta})/\nu$. The even simpler linear profile equation 
\begin{equation}
\deriv{\ }{\zeta}\left(\frac{1}{\zeta}\deriv{\chi}{\zeta}\right) - \left(1+\frac{\sigma}{\zeta}\right)\,\chi=0     \label{RAT}
\end{equation}
is recovered in the limit $\nu\to0$ with $\sigma$ held fixed (and therefore $\sin\theta\to0$). Unfortunately, neither equation admits exact solution in terms of special functions (neither \emph{Mathematica} 7 nor Maple 12 can solve them, nor do they appear in the compendious tome of \opencite{pz}). We shall term Equation (\ref{RAT}) the Reflection-Absorption/Transmission (RAT) equation, and Equation (\ref{ERAT}) the Exponential Reflection-Absorption/Transmission (ERAT) equation. Notwithstanding the above protestations about the realism of the linear profile, the RAT equation is useful here as the $\theta\to0$, $\kperp\to\infty$, $\sigma$ finite limit of the ERAT equation.

\begin{figure}[tbp]
\begin{center}
\includegraphics[width=0.59\textwidth]{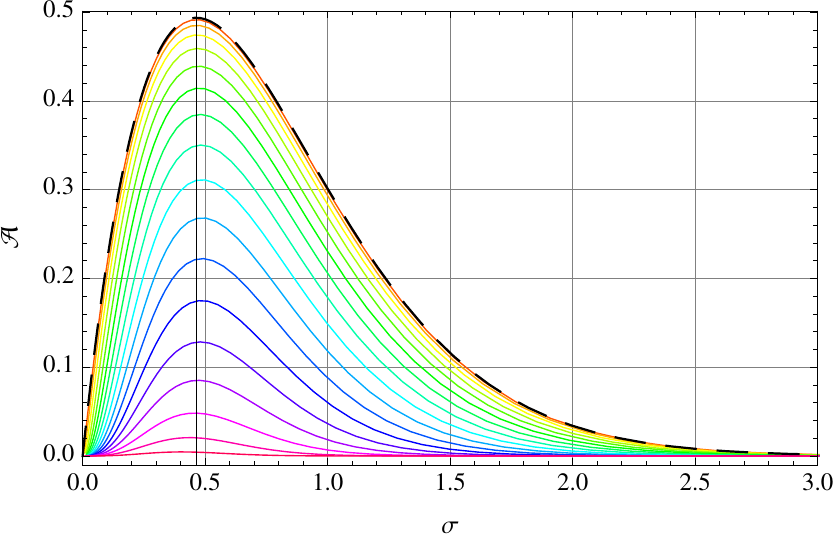}\\[6pt]
\includegraphics[width=0.6\textwidth]{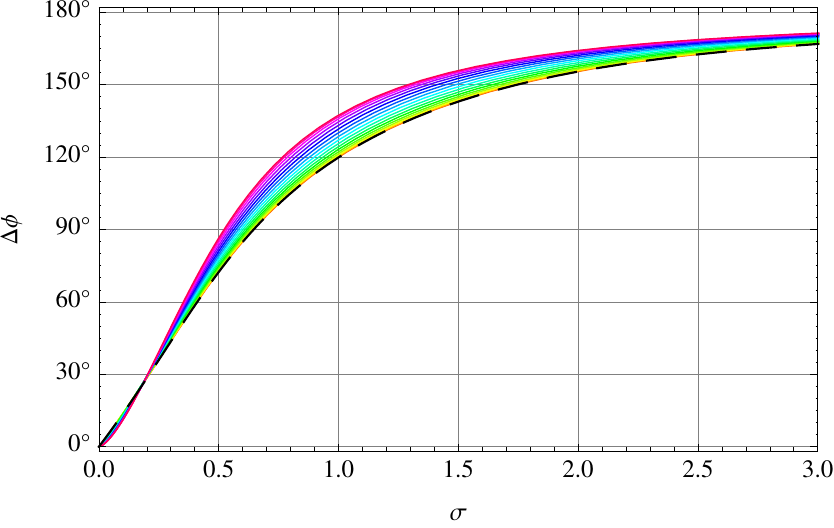}{\ }
\caption{Top panel: Absorption coefficient $\mathscr{A}$ as a function of $\sigma=\epsilon^{-2/3}h^{2/3}\sin^2\theta=(\kperp h)^{2/3}\sin^2\theta$ for $\Omega^2=e^{-x/h}$ and $\theta=5^\circ$, $10^\circ$, \ldots, $85^\circ$ (top to bottom). The heavy dashed curve is the RAT index (\emph{i.e.}, $\theta=0^\circ$). The heavy vertical line is at $\sigma=\sigma_0$. Bottom panel: Phase retardation $\Delta\phi$ between incident and reflected waves for the same cases, with corresponding colours, \emph{i.e.}, $\theta=0^\circ$, $5^\circ$,  \ldots, $85^\circ$ (bottom to top).}
\label{fig:Asig}
\end{center}
\end{figure}

Figure \ref{fig:Asig} summarizes the results of a numerical survey of the solutions of the RAT and ERAT equations with $0<\sigma<3$ and $\theta=0^\circ$ (RAT), $5^\circ$, \ldots, $85^\circ$ (ERAT; see the appendices for details). Both the total absorption coefficient $\mathscr{A}$ and the phase retardation of the reflected wave $\Delta\phi$ are plotted. They are also tabulated in Tables \ref{Asigmatheta} and \ref{Deltaphi}. Strikingly, absorption never exceeds 50\%, and vanishes as $\sigma\to0$ and $\sigma\to\infty$. These absorption curves are similar to those obtained by \inlinecite{holl90} for a two-piece linear-constant Alfv\'en profile and a nearby ``vacuum-boundary''.

\begin{figure}[htbp]
\begin{center}
\includegraphics[width=0.49\textwidth]{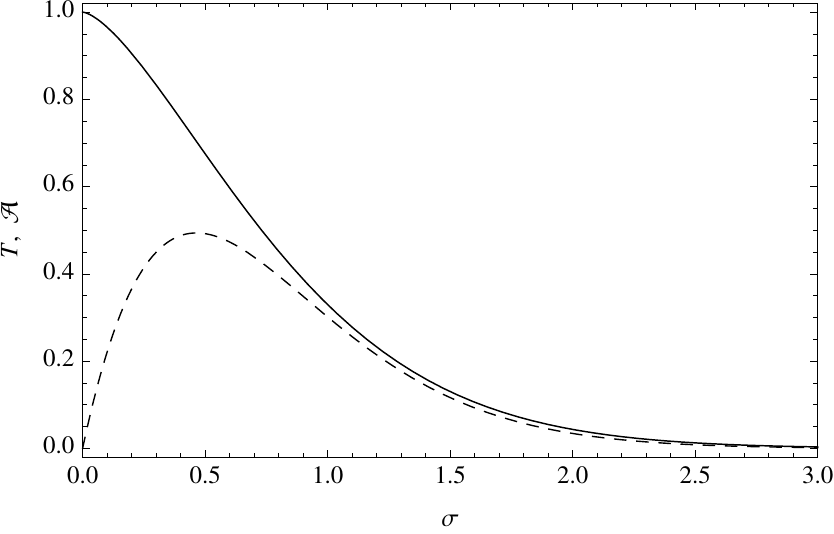}\hfill
\includegraphics[width=0.49\textwidth]{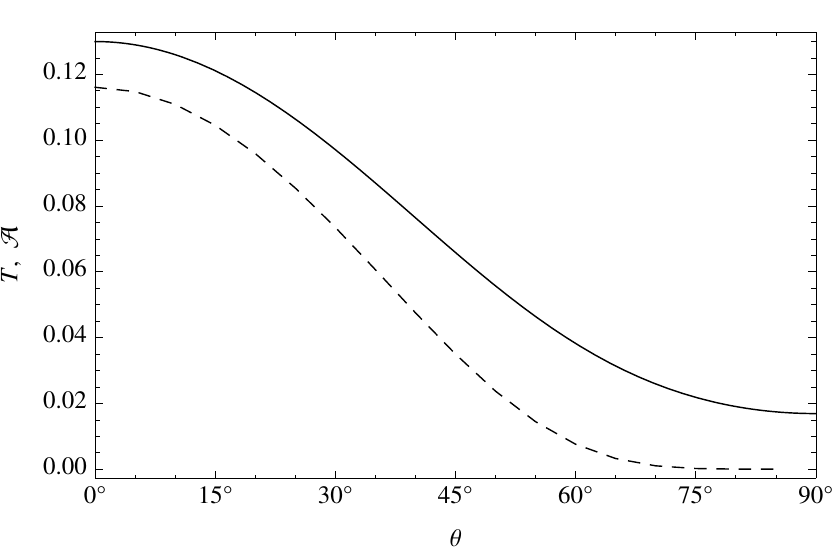}
\caption{Left: Transmission coefficient $T$ as given by Equation (\ref{T}) for $\theta=0^\circ$ and $0<\sigma<3$ (full curve) compared to the numerically calculated absorption coefficient $\mathscr{A}(\sigma,0)$ (dashed curve). Right: $T$ and $\mathscr{A}(\sigma,\theta)$ for $\sigma=1.5$ and $0^\circ<\theta<90^\circ$.}
\label{fig:TA}
\end{center}
\end{figure}

The modular theory transmission $T$ is plotted against $\sigma$ and $\theta$ in Figure \ref{fig:TA} and compared with the numerically derived values of $\mathscr{A}$. It is obviously totally wrong at small $\sigma$, though broadly correct (in absolute value and shape) at larger $\sigma$, both in terms of $\sigma$ and $\theta$ dependence. Significantly though, it is always an upper bound, $T>\mathscr{A}$.


The limited applicability of Equation (\ref{T}) is no refutation of the mode conversion concept \emph{per se}, merely a reflection of the limitations of the basic modular theory for this difficult problem involving a resonance. Nor does it render infeasible calculation based on conversion theory, since we have in fact determined \emph{exact} (numerical) values for the absorption coefficient $\mathscr{A}(\sigma,\theta)$ for the exponential profile, albeit in graphical and tabular form. Interpolation of Table \ref{Asigmatheta} is as good as having an exact formula.

\subsection{Phase}
\begin{figure}[htbp]
\begin{center}
\includegraphics{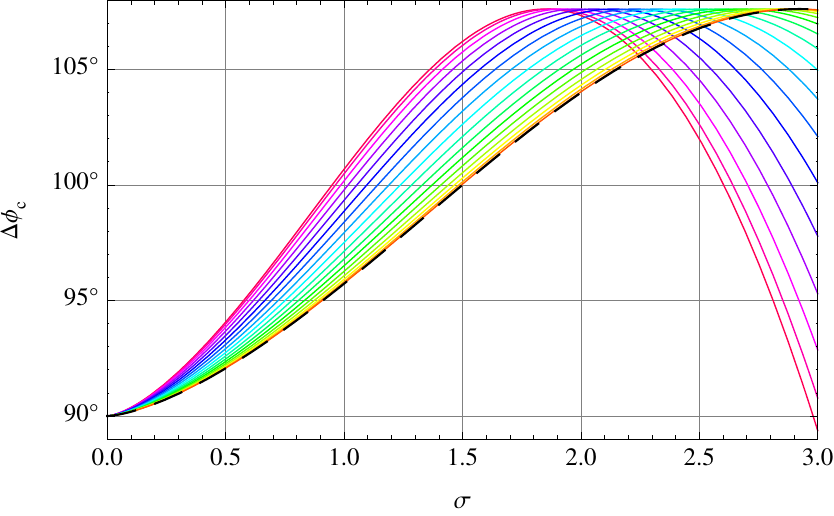}
\caption{Phase retardation due to mode conversion as predicted by Equation (\ref{cphase}). The line styles correspond to those of Figure \ref{fig:Asig}, \emph{i.e.}, $\theta=0^\circ$, $5^\circ$, $10^\circ$, \ldots, $85^\circ$ (bottom to top on $\sigma\lesssim1.8$), with the heavy dashed curve corresponding to the RAT case $\theta=0^\circ$.}
\label{fig:phic}
\end{center}
\end{figure}

Turning now to the reflected wave (the ``converted'' wave in the terminology of \opencite{tkb03}), the ``wave conversion coefficient'' 
\begin{equation}
c = \frac{\sqrt{2\,\pi\,\tau}}{\eta\,\Gamma(-\ri\,|\eta|^2)}    \label{cphase}
\end{equation}
is such that the (wave-energy) conversion coefficient $C=|c|^2$ is exactly $1-T$. However, $c$, which is complex, also contains information about the phase change from the incident to the converted wave: the phase retardation due to mode conversion is $\Delta\phi_{\rm c} = -\arg c$. This is plotted in Figure \ref{fig:phic}, and disappointingly bears little resemblance to $\Delta\phi$ as depicted in Figure \ref{fig:Asig}. This is probably because of the confounding effect of the additional reflection component to that accounted for by $T$.

Phase change is of relevance to seismology, especially as regards the determination of wave travel times \cite{cally09a,cally09b}.

\section{Superposition and Ray Bundles}   \label{bundle}
We postulate that ``trappedness'' of the resonance is fundamental to our interpretation of the physical process in operation: is it mode conversion or is it resonant absorption? Our current model does not at first sight allow us to explore the consequences of opening the resonance, because the resonant field line is being excited uniformly along its entire length. But this $z$-invariance can be avoided by inverting the $z$-Fourier transform to construct a discrete ray bundle that meets the resonance only in a limited region.

Before synthesizing a ray in this way, we should first define an absolute $x$-scale. Previously, the $x$ axis has been translated so that the reflection point is at $x=0$ for each mode. But this reflection point position depends on $k_z$. For this calculation, the zero of $x$ shall be arbitrarily placed at the reflection point for a specified $k_{z0}$ (the central wavenumber of a distribution); for other $k_z$ it is at $x=\Delta_{k_z}$, where $\Delta_{k_z}=\ln(\sin^2\theta/\sin^2\theta_0)$ for our standard exponential density profile with $h=1$, and $\theta=\arctan(k_y/k_z)$. Similarly, the resonance is shifted to $X(k_z)=\ln\sec^2\theta+\Delta_{k_z}$. In this section we denote the normal mode solution by $\hat\chi(x,k_z)$, equivalent to $\chi(x-\Delta_{k_z},k_z)$ of the previous sections, and reserve $\chi$ for the physical space solution, though the $t$ and $y$ Fourier transforms are still in operation. Indeed, we still fix a single $k_y$ and $\omega$, but now integrate over $k_z$. 
With this in mind, the physical space dilatation $\chi$ is
\begin{equation}
\chi(x,z) = \frac{1}{2\pi} \int_{-\infty}^\infty e^{\ri\,k_z z}A(k_z)\,\hat\chi(x,k_z)\, dk_z\, .   \label{invF}
\end{equation}
In Equation (\ref{invF}), $A(k_z)$ is a compact amplitude which shall be centred on a particular $k_z=k_{z0}$. If it is a delta-function, we recover a single mode, but by broadening the profile in Fourier space we sharpen it in $z$-space. A ray bundle may therefore be constructed that impacts the (spread) resonance layer and may mode convert to an Alfv\'en wave that \emph{can} escape from the interaction region by propagating in the $z$-direction. 

To this end, a gaussian amplitude distribution
\begin{equation}
A(k_z) =
\sqrt{2 \pi } \,H \exp \left[-
   \half\left(k_z-k_{z0}\right)^2H^2
   -\ri\left(k_z-k_{z0}\right)
   z_0\right]      \label{Amp}
\end{equation}
is adopted that has inverse Fourier Transform $\exp[-(z-z_0)^2/2 H^2+\ri\, z\, k_{z0}]$. This is centred at $k_{z0}$ in Fourier space and $z_0$ in physical space, and has widths $H^{-1}$ and $H$ respectively. The incoming component of $\hat\chi$ is normalized to 1 at the left boundary $x=x_\text{L}$.

The inverse transform is obtained numerically using the fast discrete method of \inlinecite{bs94} based on the Fractional Fourier Transform (FRFT). A large number $m$ of points (typically a power of two or other product of small primes suitable for FFT evaluation) is used in both Fourier and physical space lying uniformly in $[k_{z0}-K,\,k_{z0}+K)$ and $[-Z,\,Z)$ respectively, where $K$ is large enough that $A(k_z)$ is effectively zero outside the interval and $Z$ similarly encompasses the ray bundle in physical space. The simple piecewise-constant integration rule is implicit in the inversion as applied here. Each of $\hat\chi$, $\partial\hat\chi/\partial x$, $\partial\hat\chi/\partial z$, and $\partial^2\hat\chi/\partial x\partial z$ is evaluated in Fourier space and then mapped to physical space by Fourier inversion in order to construct the required displacements and fluxes.

Exact integration of the ray equations in the exponential atmosphere $\Omega^2=e^{-x}$ yields the ray path $x=2\ln\cos[\half z\sec\theta]$ for a ray passing through the origin. In particular, the ``skip distance'' in the $z$-direction for this ray is $\Delta z=2\pi\cos\theta$, which allows us to estimate \emph{a priori} the $z$-range required in the numerical Fourier inversion.

The following parameters are adopted by way of example: $m=2^8=256$, $k_y=3$, $k_{z0}=9$, $K=8$, $H=0.5$, $x_\text{L}=-5$, $z_0=-\pi$, and $Z=5$. For this case, $0.2<\sigma<1.94$ and $10.0^\circ<\theta< 71.6^\circ$, with $\sigma=0.448$, $\theta=18.43^\circ$, $X=0.105$, and $\mathscr{\hat A}=0.464$ at $k_z=k_{z0}$, suggesting strong absorption. Figure \ref{fig:ph} depicts the phase space structure for the central wavenumber. For the bundle, the resonance point $x=X(k_z)$ is now widely spread because of the distribution in $k_z$ space; for $|A(k_z)|$ greater than half-maximum it spans the interval $-0.35<x<0.7$.

\begin{figure}[htbp]
\begin{center}
\includegraphics[width=.8\textwidth]{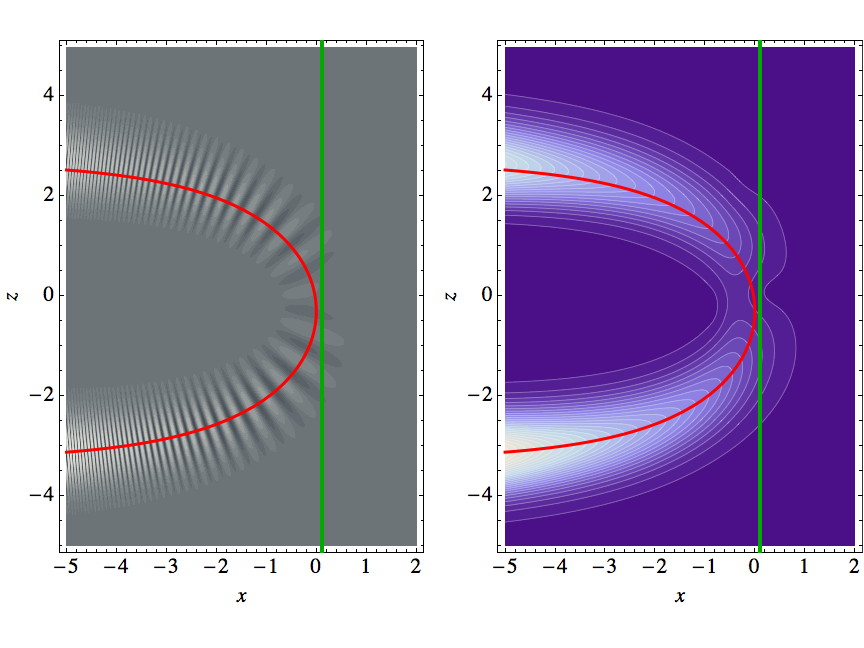}
\caption{Contours of $\re\chi$ (left) and $|\chi|$ (right) constructed by inverse Fourier Transform plotted in the $x$-$z$ plane for the example case discussed in the text of Section \ref{bundle}. The grid used is $561\times256$. The exact ray path of the central $k_z$ is over-plotted in red, and the green line indicates the position of the resonance for this central value. The ray enters the domain from the lower left boundary and leaves, with reduced intensity, via the upper left boundary. Dark purple in the right panel corresponds to 0 and maximum values to white shading. The contour levels are evenly spaced. 
}
\label{fig:contChi}
\end{center}
\end{figure}

Figure \ref{fig:contChi} displays the result of this calculation as contour plots of $\re\chi$ and $|\chi|$ over the $x$-$z$ plane. A clearly defined ray bundle is apparent which fits very well with the analytic ray path (in red). It enters around $z_0=-\pi$ on the left, reflects near $x=0$, and leaves the domain at positive $z$ on the left. This is as we might expect. But note that the intensity of the bundle is less by some 26\% on exit than on entry, indicating the energy must have gone elsewhere. 
We should rather look at $F_z = F_0 \re\{\ri\,\bxi\vdot\partial\bxi^*/\partial z\}$.

\begin{figure}[thbp]
\begin{center}
\includegraphics[width=.8\textwidth]{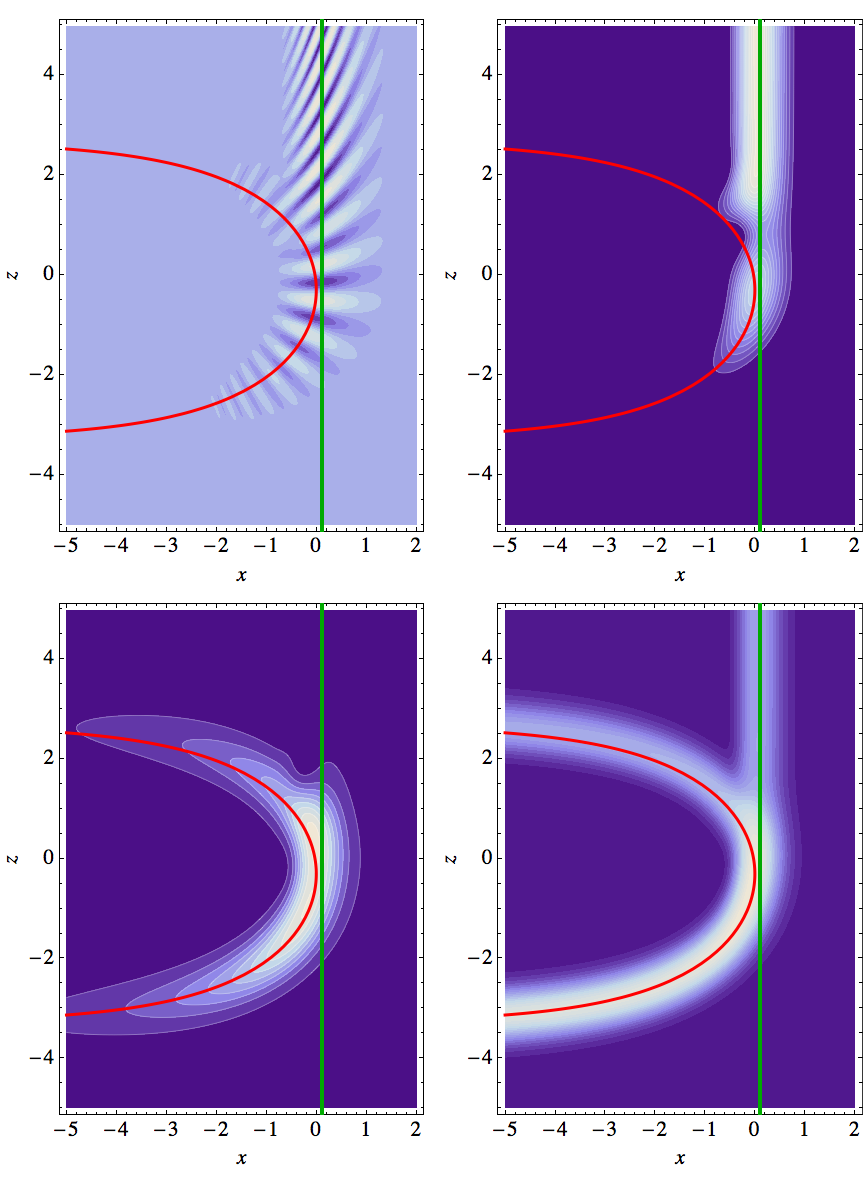}
\caption{Displacements and fluxes for the case of Figure \ref{fig:contChi}. Top left: contours of $\re\xi_y$, characteristic of the Alfv\'en wave, and displaying phase mixing with increasing $z$ along the resonance band. Top right: contours of $F_{z,\text{Alf}}$. Bottom left: $F_{z,\text{fast}}$. Bottom right: shading representing absolute $x$-$z$ wave energy flux $|F_{xz}|=(F_x^2+F_z^2)^{1/2}$. Different colour scaling ranges are used for the three panels, corresponding in each case to white being the maximum, and the broad background colour corresponding to zero. The maximum values for $F_{z,\text{Alf}}$, $F_{z,\text{fast}}$, and $|F_{xz}|$ are respectively 0.0044, 0.0072, and 0.0099. $F_0=1$ is assumed throughout.}
\label{fig:conts}
\end{center}
\end{figure}

Since $\hat\chi$ is a $\mathcal{C}^1$ function Fourier inversion results in $\chi$ being $\mathcal{C}^2$. On the other hand, the singularities in $\hat\bxi$ make direct numerical inversion problematic. It is preferable instead to construct the displacement from $\chi$ by solution of ordinary differential equations in physical-space:
\begin{equation}
\pderivd{\bxi}{z}+\frac{\omega^2}{a^2}\bxi = -\grad_{\!p}\chi\,,
\label{xiDirect}
\end{equation}
where $\grad_{\! p}$ is the $x$-$y$ part of the gradient operator. Equation (\ref{xiDirect}) is easily derived from Equations (\ref{xiDEs}), and is solved subject to $\bxi\to\mathbf{0}$ as $z\to-\infty$, yielding $\mathcal{C}^3$ solutions. The interpretation of Equation (\ref{xiDirect}) is clear: $\bxi$ propagates Alfv\'enically along the field lines with $-\grad_{\! p}\chi$ acting as a source term. The exact solution is
\begin{equation}
\begin{split}
\bxi(x,z) 
& = \frac{1}{k}\int_{-\infty}^z \!\! \grad_{\! p}\chi(x,z')\,\sin k(z'-z)\,dz'   \\
&= \frac{1}{2\,\ri\, k} \int_{-\infty}^z \!\! \grad_{\! p}\chi(x,z')\left[e^{\ri k(z'-z)} -e^{-\ri k(z'-z)}\right]\,dz'
 \, ,   
 \end{split}
 \label{varpar}
\end{equation}
where $k(x)=\omega/a=\Omega\kperpz$, and $\kperpz=\sqrt{k_y^2+k_{z0}^2}\,$. Resonant transfer of energy from $\chi$ to $\bxi$ clearly occurs where $\grad_{\! p}\chi$ has a substantial Fourier component resonant with $k$. In practice, the integrals are performed on the grid using a modified trapezoidal rule\footnote{The trigonometric form is used, with $\sin k(z'-z)$ first expanded by multiple angle formula. To ameliorate the effects of imperfect resolution of $\cos kz'$ and $\sin kz'$ on the grid $z_1$, $z_2$, \ldots, $z_m$, we apply a piecewise-linear approximation for $\grad_{\! p}\chi$ and perform the resulting trigonometric-by-linear integrals exactly.  Higher order schemes are unnecessary for our purposes.} and the lower boundary condition is applied at $z=-Z$ on the basis that there is no significant signal below that point.

Equation (\ref{varpar}) may be interpreted more generally. The sine, or the combination of exponentials, may be thought of as eigenfunctions of the Alfv\'en operator along the field lines. The displacement $\bxi$ is therefore constructed by projection of the compressional source term onto these eigenfunctions. We believe this result will carry over to more complex examples than the simple case considered here.

\begin{figure}[tbhp]
\begin{center}
\includegraphics[width=\textwidth]{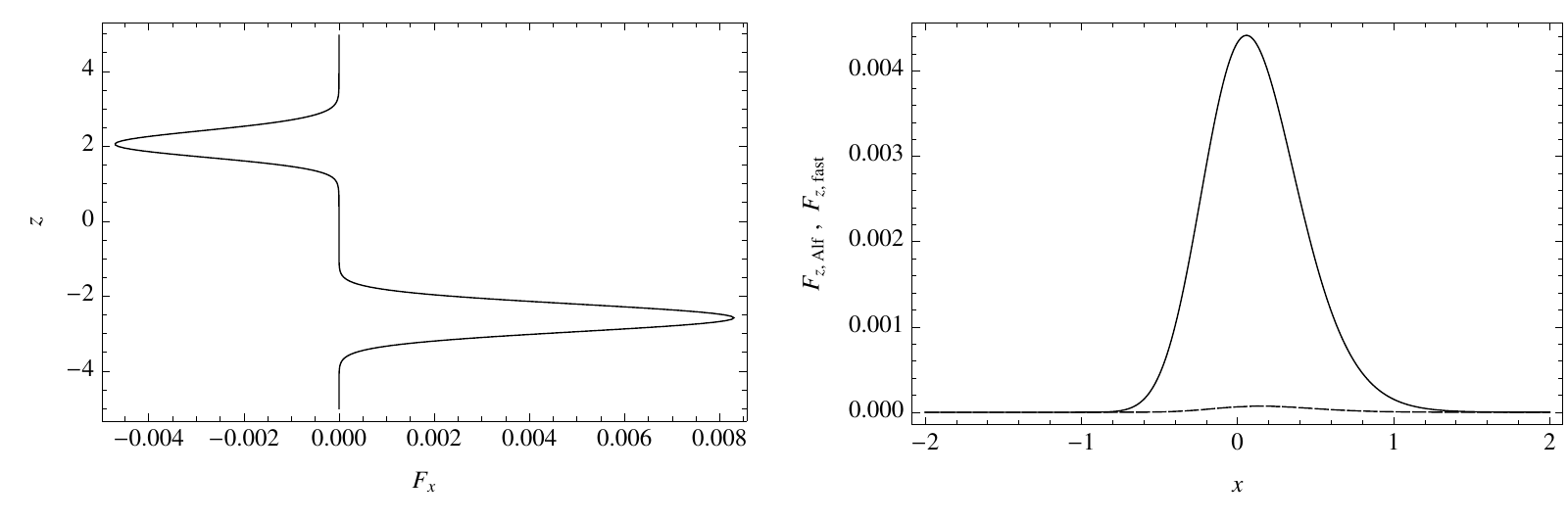}
\caption{The fluxes through $x=-2$ (left panel) and the top boundary (right panel). In the latter case, $F_{z,\text{Alf}}$ (solid) and $F_{z,\text{fast}}$ (dashed) are displayed separately.}
\label{fig:Finout}
\end{center}
\end{figure}

Though $\chi$ is a poor marker of an Alfv\'en wave, $\xi_y$ is ideal. It is displayed in the top left panel of Figure \ref{fig:conts}, making very clear the generation and propagation of an Alfv\'en wave in the resonant slab. Phase mixing is also apparent as the Alfv\'en wave propagates upwards, caused by differing propagation speeds on neighbouring field lines. Now, the $x$ and $z$ components of wave energy flux are $F_x=F_0 \re(\ri\,\chi^*\,\xi_x)$ and $F_z=F_0\,\re(\xi_x\,\partial\xi_x^*/\partial z + \xi_y\,\partial\xi_y^*/\partial z)=F_{z,\text{fast}}+F_{z,\text{Alf}}$, where $F_{z,\text{Alf}}$ is that part of $F_z$ associated with $\xi_y$. The top right panel of Figure \ref{fig:conts} again convincingly shows the presence of an Alfv\'en wave taking energy upwards and away from the interaction by exclusively plotting $F_{z,\text{Alf}}$. On the other hand, $F_{z,\text{fast}}$ (bottom left) quickly fades out in the $z$-direction. 

Finally, and most strikingly, the bottom right panel of Figure \ref{fig:conts} displays contours of the total $x$-$z$ flux, showing how it splits into the reflected and Alfv\'enic branches, each with lesser energy than the incident branch. One further interesting though subtle feature most apparent in this last graphic is that the outgoing ray has been shifted slightly upwards relative to the ray theory path. We believe this is a consequence of the phase shift associated with the mode conversion.

The reader is strongly advised to view the movie accompanying this paper as electronic supplementary material. It shows the dilatation field in greyscale (as in Figure \ref{fig:contChi}a), but with the Alfv\'en displacement $\xi_y$ overplotted in blue and red, representing displacements out of and into the page respectively. The manner in which the two fields move in phase is most instructive.

Figure \ref{fig:Finout} shows the energy balance between the incoming, reflected and Alfv\'enic ray bundles. Total incoming flux on $x=-2$, $z<0$ is 0.00761 and outgoing flux on $x=-2$, $z>0$ is 0.00417, with net flux 0.00343. The reflection coefficient is therefore 0.55, which accords very well with $1-\mathscr{\hat A}(\sigma,\theta)=0.54$ for the central wavenumber $k_{z0}$. The net outgoing flux on the top is 0.00340, matching almost perfectly (a small amount of energy is lost because of imperfect resolution on the grid and the low-order integration schemes used).

\begin{figure}[tbhp]
\begin{center}
{\hfill\includegraphics[width=.4\textwidth]{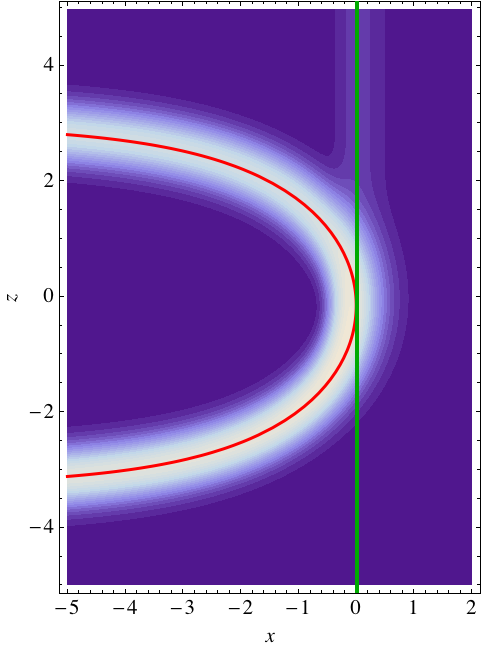}\;\;
\includegraphics[width=.4\textwidth]{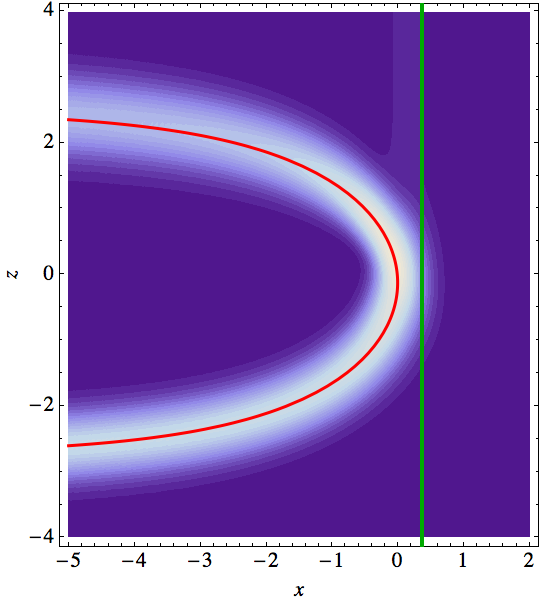}\hfill}
\caption{Same as the bottom left panel of Figure \ref{fig:conts}, except with $k_y=1$ (left) and $k_y=6$ (right).}
\label{fig:Ftotky}
\end{center}
\end{figure}

By way of comparison, Figure \ref{fig:Ftotky} shows the total flux for the cases with $k_y=1$ and $k_y=6$ and the other parameters unchanged. These examples have central $\sigma=0.053$ and $\theta=6.3^\circ$ for $k_y=1$, and $\sigma=1.505$ and $\theta=33.7^\circ$ for $k_y=6$, with corresponding narrower and wider evanescent gaps respectively and much lower central absorptions $\mathscr{\hat A}=0.12$ and $0.063$. As expected, the ray bundles suffer only very weak Alfv\'en conversion (0.12 and 0.064), and there is little apparent upward displacement of the reflected rays.

So how exactly are the reflection and absorption coefficients of a ray bundle related to those for the constituent modes? Integrating the physical-space $x$-flux $F_x(x,z)=F_0 \re(\ri\chi^*\xi_x)$ over all $z$ at fixed $x$, it is easily shown that 
\begin{equation}
\begin{split}
F_\text{tot}=
\int_{-\infty}^\infty F_x(x,z)\, dz &=  \frac{1}{2\pi} \int_{-\infty}^\infty |A(k_z)|^2\,\hat F_x(x,k_z) \, dk_z \\[4pt]
&= \langle \hat F_x\rangle\,  \frac{1}{2\pi} \int_{-\infty}^\infty |A(k_z)|^2 \, dk_z\, 
\end{split}   \label{Ftot}
\end{equation}
(defining $\langle\ldots\rangle$), where $\hat F_x$ is the flux in an individual mode $k_z$, provided the integrals exist (which is so for our gaussian amplitude). For the gaussian profile, $\frac{1}{2\pi} \int_{-\infty}^\infty |A(k_z)|^2 \, dk_z=H\sqrt{\pi}$. If we now split $F_x$ into its rightward and leftward components at $x_L$, $F_\text{tot}=F_x^+ + F_x^-$, and similarly for $\hat F_x$, it is readily found that
\begin{equation}
\mathscr{R} = -\frac{F_x^-}{F_x^+}=\frac{\langle {{\mathscr{\hat R}}{\hat F_x^+}} \rangle}{\langle{\hat F_x^+}\rangle}\,,  \label{Rfourier}
\end{equation}
where $\mathscr{R}$ is the overall reflection coefficient of the ray bundle and $\mathscr{\hat R}$ is that for the individual Fourier modes. More simply, if $\hat F_x^+$ is scaled to unity, then $\mathscr{R}=\langle\mathscr{\hat R}\rangle$. This shows how the reflection coefficient for a ray bundle may be constructed from those for the constituent modes. Clearly, the sharper is $A$ in $k_z$ space, the more closely $\mathscr{R}$ matches $\mathscr{\hat R}$. But this corresponds to a very broad ray bundle in physical space. If on the other hand we sharpen the bundle in $z$, it is broadened in $k_z$, resulting on a broad flux-weighted average over $k_z$. 


Selecting $z$ well above the region where $\chi$ is significant, Equation (\ref{varpar}) may be integrated exactly to determine the Alfv\'enic flux at the top $F_{z,\text{top}}(x)=F_{z,\text{Alf}}(x,\infty)$ on each field line. Curiously, this physical-space quantity depends algebraically on the Fourier-space dilatation $\hat\chi$:
\begin{equation}
F_{z,\text{top}}(x)=
\frac{F_0\, k_y^2}{4\,k(x)} \, \left|A\left(k(x)\right)\right|^2 \,
 \left|\hat\chi\left(x,k(x)\right)\right|^2\, .     \label{Ftop}
\end{equation}
It is peaked close to where $A(k(x))$ is peaked, \emph{i.e.}, at $k=\omega/a=k_{z0}$, the 
central resonant point, consistent with Figures \ref{fig:conts} and \ref{fig:Finout}. This expression for $F_{z,\text{top}}(x)$ is exact and does not suffer the slight numerical inaccuracies associated with the integrations for $\bxi$.

Equation (\ref{Ftot}) and Equation (\ref{Fx}) for $\hat F_x$ may be combined to show
\begin{equation}
\begin{split}
F_\text{tot}(x_2) -F_\text{tot}(x_1) &=\int_{-\infty}^\infty F_x(x_2,z)\, dz - \int_{-\infty}^\infty F_x(x_1,z)\, dz \\
&=  \frac{1}{2\pi} \int_{-\infty}^\infty |A(k_z)|^2\, \left(\hat F_x(x_2,k_z) - \hat F_x(x_1,k_z) \right)\,dk_z\\
&= F_0\, k_y^2 \int_{x_1}^{x_2} \frac{|A(k(x))|^2}{4\,k(x)}\, |\hat\chi(x,k(x))|^2\, dx
= \int_{x_1}^{x_2} F_{z,\text{top}}\, dx\,,
\end{split}
\end{equation}
where $x_2>x_1$ and $dk/dx=-k/2h$ has been used. Letting $x_2\to x_1^+$ confirms that indeed the energy entering and being consumed by the resonance at $x_1$ is exactly that converted to $F_z$ along that field line.

The significance of Equation (\ref{Ftop}) is that the Alfv\'enic flux is made up of \emph{precisely} the modal dilatation on the resonances; at each point $x$ the flux comes solely from the mode which has its resonance there. And providing $A(k)$ is smooth, then so is $F_{z,\text{top}}$, despite $\hat\bxi$ being undefined at the resonance. This is a pretty insight.

\section{Discussion and Conclusions}    
The analysis of mode transmission and absorption presented here for the simple case of a cold MHD plasma allows us to at last understand the commonalities and distinctions between mode conversion and resonant absorption theories. 

Both the current model and the mode conversion models of \inlinecite{cally06} and \inlinecite{sc06} share the characteristic that there is an avoided crossing or near passage of two distinct branches of the dispersion relation somewhere in $\x$-$\k$ phase space, across which energy may tunnel. In the case of the \opencite{sc06} analysis, an upward-travelling fast MHD wave with predominantly acoustic characteristics (since the sound speed $c$ exceeds the Alfv\'en speed $a$ in the solar interior) is incident on a surface active region where $a\approx c$ and transmits to a slow (but still predominantly acoustic) wave in the region above ($a>c$). The field-guided sound wave then takes the energy away to be either deposited in higher layers or to reflect from an acoustic cutoff.

On the other hand, the modal resonant absorption model discussed here provides no opportunity for energy to be removed from the interaction region. Although a singular Alfv\'en wave of sorts propagates along the $z$-directed field line at $x=X$, there is no net Alfv\'enic energy loss from any segment. Instead, energy is being continually stored deeper and deeper in the singularity (recall that it takes an infinite time for energy to actually reach $X$), or in reality is thermalized there by non-ideal effects. 

In particular, we may now understand resonant absorption as something that occurs when there is transmission to a ``trapped'' mode. This consideration deserves some further elaboration as it can easily be misinterpreted in various ways.

First, one may have the impression that the resonances are related to the ``closedness'' of the system which artificially traps the energy at a certain position. This is strikingly untrue. In fact, in the modal model of Section \ref{sec:model} the field lines are open and hence an Alfv\'enic wave packet can propagate away in the $z$ direction, though with no net Alfv\'enic loss, and yet a singularity still results. This brings us to the second possible misinterpretation, previously expressed by some authors \cite{schwartz1984,hansen-harrold1994}, that the occurrence of resonances and the associated singularities is due to studying a too simple model (see \opencite{rw99}, for a slightly different slant on the \opencite{hansen-harrold1994}, results). A well-known example is the case of horizontal magnetic field in a gravitationally stratified atmosphere (\emph{e.g.}, \opencite{cally84}), where the linear wave equations reduce to a single second order ordinary differential equation with coefficient of the second derivative vanishing at Alfv\'en and slow resonances thereby leading to classic resonant absorption. However, if the field is tilted only slightly, the equations become fourth order and the resonance is gone! The reason is that there has been mode conversion to a wave that can run upward along the field line to escape, whereas in the horizontal field case it has nowhere to go. This example, however, is misleading. The horizontal field case suffers the same shortcoming as our present model, \emph{i.e.}, the invariance along the field. The inclined field case seems to resolve that issue but at the cost of simplifying the model in another respect. In the case of an inclined field all field lines are alike and the propagation across the field is trivialised (by Fourier expansion). It is exactly this aspect that removes the resonances \cite{goossens1985} which are essentially associated with the non-trivial behaviour of solutions across the field. Resonant absorption has indeed been studied in models including both variation along and across the field simultaneously (\emph{e.g.}, \opencite{thompson1993}). 

Based on these considerations it is clearly necessary to remove the invariance along field lines in order to progress further with our understanding. This is done in Section \ref{bundle} by synthesizing a discrete ray bundle from the previously calculated singular normal modes. The resulting ray is not only smooth in $\chi$ as expected, but also smooth in the displacement $\bxi$ as constructed directly from $\chi$. There is no sign in physical space of a local or even distributed energy build-up, and also the escaping Alfv\'en wave is strikingly apparent. This Alfv\'en wave excitation/conversion mediated by resonant absorption has been noted previously for kink waves in coronal loop models \cite{rr02}.

So is it ``resonant absorption as mode conversion"? Or is it ``mode conversion as resonant absorption"? It is both! We have fabricated a (mode converting) ray bundle from the (resonant absorbing) normal modes, but could just as well have constructed the normal modes from a collection of rays.

The fundamental and inescapable conclusion therefore is that resonant absorption and mode conversion are indeed two aspects of the same process. The distinguishing characteristic is whether or not the system allows a \emph{net} escape of energy from the absorbing region in physical space. If so, it is best understood in terms of mode conversion; if not, then energy escapes in wavenumber space only, producing arbitrarily small scales (or local non-ideal dissipation), and resonant absorption theory is the most natural description.

\begin{acknowledgements}
The authors wish to acknowledge helpful discussions with Gene Tracy, Alain Brizard, and Yanli Xiao. JA is supported by an International Outgoing Marie Curie Fellowship within the 7th European Community Framework Programme. JA acknowledges support by the National Fund for Scientific Research -- Flanders (Belgium) (FWO-Vlaanderen).
\end{acknowledgements}


\appendix
\section{Numerical Solution for a General Profile}   \label{app:gen}

\subsection{Eikonal Description}  \label{sub:eik}

We follow the general asymptotic WKB expansion method as described by \inlinecite{bo}, using $\epsilon=\kperp^{-1}$ as our ``small parameter'':
\begin{equation}
\qquad\qquad\qquad\chi \sim \exp\left[\sum_{n=0}^\infty \epsilon^{n-1}S_n\right]   \, , \qquad\qquad \epsilon\to0\,, \label{WKBexp}
\end{equation}
and obtain a WKB hierarchy for $S_0$, $S_1$, etc. Defining $K^2=\beta\cos^2\theta-\sin^2\theta=(\beta-\beta_0)\cos^2\theta=\Omega^2-1$, 
\begin{subequations}\label{WKBhier}
\begin{align}
S_0'^2 &= -K^2\label{WKB0}\\[4pt]
2S_0'S_1'+S_0''-(\beta'/\beta)S_0' &= 0 \label{WKB1}\\[4pt]
S_1'^2+2S_0'S_2'+S_1''-(\beta'/\beta)S_1' &= 0 \label{WKB2}
\end{align}
\end{subequations}
etc., whence 
\begin{subequations}\label{S}
\begin{align}
S_0 &= \int^x\sqrt{-K(t)^2}\,dt\label{S0}\\
S_1 &= \ln\left[(-\beta)^\frac{1}{2}(-K^2)^{-\frac{1}{4}}\right] \,.\label{S1}
\end{align}
\end{subequations}
Consequently, the physical optics approximation (\emph{i.e.}, up to and including $S_1$) is
\begin{equation}
\chi \sim (-\beta)^\frac{1}{2} (-K^2)^{-\frac{1}{4}} \left\{ A \exp\left[\frac{1}{\epsilon}\int^x\sqrt{-K(t)^2}\,dt\right] + B \exp\left[-\frac{1}{\epsilon}\int^x\sqrt{-K(t)^2}\,dt\right]\right\}  \label{chigen}
\end{equation}
as $\epsilon\to0$, where $A$ and $B$ are integration constants. The validity of this approximation requires that $\epsilon S_2 \ll 1$ as $\epsilon\to0$. In Region I, where both $\beta$ and $K^2$ are positive, an alternate normalization will be used and $S_0$ becomes imaginary, 
\begin{equation}
\chi \sim \beta^\frac{1}{2} (K^2)^{-\frac{1}{4}} \left\{ A_1 \exp\left[\frac{\ri}{\epsilon}\int^x\sqrt{K(t)^2}\,dt\right] + B_1 \exp\left[-\frac{\ri}{\epsilon}\int^x\sqrt{K(t)^2}\,dt\right]\right\} \, . \label{chigenI}
\end{equation} 
The two complex exponentials are respectively the incident and the reflected waves.

\subsection{Frobenius Solution}    \label{sub:frob}
For the case of a general monotonic profile, Equation (\ref{DE}) is solved numerically using an initial value ODE solver starting from the decaying eikonal approximation (up to and including $S_2$) at $x\gg X$, \emph{i.e.}, with $A=0$ and $B=1$ in (\ref{chigen}), and integrating leftward. However, to avoid integrating through the singularity at $x=X$, the Frobenius solution (below) is used in a neighbourhood $(X-\delta,X+\delta)$. The numerical solution is then continued from $X-\delta$ to a large negative $x$ where the eikonal solution (\ref{chigen}) may again be used to identify the incoming ($A$) and outgoing ($B$) parts.

To develop the Frobenius solution about the resonance point assume that $\beta$ is analytic in a neighbourhood of $X$, and write
\begin{equation}
\beta=\sum_{n=1}^\infty\beta_n(x-X)^n \, ,     \label{beta}
\end{equation}
where $\beta_n=\beta^{(n)}(X)/n!$ and $\beta_1<0$. The dependent variable $\chi$ may then be expanded in a Frobenius series for which the indicial equation yields $r=2$ and 0. The first solution is
\begin{equation}
\chi_a = \sum_{n=0}^\infty a_n(x-X)^{n+2}\, ,    \label{chia}
\end{equation}
whilst the second is
\begin{equation}
\chi_b = \chi_a\ln (x-X) + \sum_{n=0}^\infty A_n(x-X)^{n}\, .    \label{chib}
\end{equation}
The causal branch of the logarithm in this instance is the one where $\ln(x-X)=\ln(X-x)-\ri\pi$ on $x<X$.

The coefficients are $a_0=1$, and
\begin{equation}
a_{p-1} = \frac{\epsilon^{-2}\left(D_p\sin^2\theta-C_p\cos^2\theta\right)-E_p}{(p^2-1)\,\beta_1}\, ,
\end{equation}
where
\begin{equation}
\begin{aligned}
C_p &= \sum_{n=0}^{p-4}a_n\gamma_{p-n-2}&
\gamma_q &= \sum_{k=1}^{q-1}\beta_k\beta_{q-k}\\
D_p &= \sum_{n=0}^{p-3}a_n\beta_{p-n-2}&
E_p &= \sum_{n=0}^{p-2}(n+2)(2n+1-p)a_n\beta{p-n} 
\end{aligned}
\end{equation}
for $p=2$, 3, \ldots. The convention is adopted that a sum is zero if its upper limit is less than its lower limit. Thus for example $D_2=0$.

Similarly,
\begin{equation}
A_0 = -\frac{\epsilon^2 Q_1}{\beta_1\sin^2\theta}=\frac{2\epsilon^2}{\sin^2\theta}\, ,
\end{equation}
$A_1=A_2=0$, and
\begin{equation}
A_{p+1} = \frac{Q_p+G_p+\epsilon^{-2}\left(J_p\sin^2\theta-L_p\cos^2\theta\right)}{(p^2-1)\,\beta_1}\, ,
\end{equation}
$p=2$, 3, \ldots, where
\begin{equation}
\begin{aligned}
Q_p &= \sum_{n=0}^{p-1} (p-3n-3)a_n \beta_{p-n}& 
G_p &= \sum_{n=1}^p n(p-2n+3)A_n\beta_{p-n+2}\\
J_p &= \sum_{n=0}^{p-1}A_n\beta_{p-n}&
L_p &= \sum_{n=0}^{p-1}A_n\gamma_{p-n} \, .
\end{aligned}
\end{equation}

With these formulae, the $a_n$ and $A_n$ coefficients may be built up recursively to any desired level. The leading terms of this expansion for the two linearly independent solutions are
\begin{equation}
\begin{aligned}
\chi_a &= (x-X)^2+\frac{2\beta_2}{3\beta_1}(x-X)^3+\frac{1}{8} \left(\frac{4 \beta _3}{\beta _1}+\frac{\sin^2\theta}{\epsilon ^2}\right)(x-X)^4+\mathcal{O}\left((x-X)^5\right)\,,\\
\chi_b &= \chi_a\ln(x-X) + 2\epsilon^2\csc^2\theta-\left(\frac{2 \beta_1 }{3}\cot^2\theta +\frac{5 \beta_2}{9 \beta_1}\right)(x-X)^3+\mathcal{O}\left((x-X)^4\right)  \, .
\end{aligned}
 \label{froblead}
\end{equation}
The logarithmic term is responsible for the absorption. 

\subsection{Reflection and Absorption Coefficients and Phase Retardation}
The overall effect of on an incident fast wave in Region I may be characterized by the reflection coefficient $\mathscr{R}=|B_1/A_1|^2$ and the associated absorption coefficient $\mathscr{A}=1-\mathscr{R}$, as well as by the net phase retardation $\Delta\phi=\arg(A_1B_1)$. This may be achieved numerically for any chosen Alfv\'en profile $\beta$, but is primarily dependent on its local properties around the reflection/absorption region. 

It remains to justify the above expression for $\Delta\phi$. If $\chi_{\rm inc}\sim e^S=e^{S_r}e^{\ri S_i}$ is the $x\to-\infty$ asymptotics for the canonical incident wave and $\chi_{\rm ref}\sim e^{S^\star}=e^{S_r}e^{-\ri S_i}$ that for the reflected wave, and $\chi = A\,\chi_{\rm inc} + B\,\chi_{\rm ref}$ is the combined solution which matches to exponential decay as $x\to\infty$, then
$
\chi \sim
e^{S_r} \left[\, |A| e^{\ri\left(S_i + \phi_A\right)} + |B| e^{-\ri\left(S_i + (-\phi_B)\right)}\right] 
$,
where $\phi_A=\arg A$ and $\phi_B = \arg B$. The incident wave is therefore advanced in phase by $\phi_A$ and the reflected wave is retarded by $\phi_B$.
The total phase retardation of the reflected wave with respect to the incident wave is therefore $\Delta\phi(\sigma,\theta)=\phi_A + \phi_B=\arg AB$.

\section{Linear Profile $\Omega^2=1-x/h$: The RAT Index}  \label{app:RAT}

Consider the case of the linear profile $\Omega^2=1-x/h$. Letting $\zeta=\epsilon^{-2/3}h^{-1/3} (x-X)$ and $\sigma=\epsilon^{-2/3}h^{2/3}\sin^2\theta$, Equation (\ref{DE}) takes the form (\ref{RAT}).
Subject to the boundary condition $\chi\to0$ at $\zeta\to\infty$ and an arbitrary linear scaling, it has a unique solution $\chi=\psi_\sigma$ for each value of the parameter $\sigma$, and consequently a unique absorption coefficient, the RAT index, $\mathscr{A}(\sigma)$ which can be calculated numerically; it is tabulated in Table \ref{Asigmatheta} as the $\theta=0$ column, displaying a peak of $A_0=0.493698$ at $\sigma=\sigma_0=0.464434$. Absorption is very well fitted by $\mathscr{A}(\sigma)\doteqdot 2.66904\,\sigma \exp[-0.363841 \sigma ^2-1.81607 \sigma] $, with maximum absolute error of 0.001 over $0\le\sigma\le3$.

A Frobenius expansion about the regular singular point $\zeta=0$ returns the two linearly independent solutions
\begin{equation}
\chi_1 = \sum_{n=0}^\infty c_n \zeta^{n+2}, \qquad \chi_2=\sigma\,\chi_1\ln\zeta+\sum_{n=0}^\infty C_n \zeta^n\, ,   \label{RATfrob}
\end{equation}
where negative $\zeta$ should be identified with $|\zeta|e^{-\ri\pi}$. The coefficients are given by the recurrence formulae
\begin{equation}
c_0=1, \quad c_1=0, \quad c_2=\frac{\sigma}{8}, \quad c_n=\frac{c_{n-3}+\sigma\, c_{n-2}}{n(n+2)},   \label{RATfrobc}
\end{equation}
and
\begin{equation}
C_0=2, \quad C_1=C_2=0, \quad C_n=\frac{\sigma 
   C_{n-2}+C_{n-3}-2 \sigma (n-1) c(n-2)}{(n-2) n}\, .  \label{RATfrobC}
\end{equation}
In the case $\sigma=0$ the logarithmic term in $\chi_2$ vanishes, leaving only Airy solutions $\Ai'(\zeta)$ and $\Bi'(\zeta)$, and indeed we expect the general solutions to resemble these in some sense for $\zeta\gg\sigma$. 

As an alternative to the WKB solutions at large $|x|$, a local analysis about the irregular singular point $\zeta=\infty$ may be used, revealing that the two linearly independent solutions $\psi(\sigma;\zeta)$ and $\phi(\sigma;\zeta)$ have asymptotic behaviours
\begin{multline}
 \zeta^{1/4}\exp\left[\pm\left(\frac{2}{3}\zeta^{3/2}+\sigma \zeta^{1/2}\right)\right] \\
\times\left(1\mp\frac{1}{4}\sigma^2\zeta^{-1/2}+\frac{\sigma ^4-8
   \sigma }{32} \zeta^{-1}
   \mp\frac{\sigma ^6-40 \sigma ^3-56}{384}
   \zeta ^{-3/2}+\cdots\right)   \label{asym}
\end{multline}
as $\zeta\to\infty$. We define $\psi(\sigma;\zeta)$ to be the ``physical'' solution for which $\psi\to0$ as $\zeta\to+\infty$ along the positive real axis, \emph{i.e.}, 
\begin{equation}
\psi\sim \zeta^{1/4}\exp\left[-\left(\frac{2}{3}\zeta^{3/2}+\sigma\zeta^{1/2}\right)\right]  \label{decay}
\end{equation}
for $\zeta>0$. Note that, although they share the same contolling factor, $\psi(\sigma;\zeta)\not\sim\psi(0;\zeta)$ as $\zeta\to\infty$, so $\sigma$ has a profound effect even at large $\zeta$. For $\sigma=0$, (\ref{decay}) is just the asymptotic representation of $-2\sqrt{\pi}\Ai'(\zeta)$.


\section{Exponential Profile $\Omega^2=e^{-x/h}$}
If Region III is sufficiently narrow ($X$ small, \emph{i.e.}, $\theta$ small) it suffices to replace $\beta$ by its tangent there, but in practice this produces satisfactory accuracy only for $|\theta|\lesssim15^\circ$. A better model is to replace $h = -(\rd\log\Omega^2/\rd x)^{-1}$ by its value at the resonance, \emph{i.e.}, to set $\Omega^2=e^{-x/h}$ with constant $h$. Then $\beta=e^{-(x-X)/h}-1$, $X=h\ln\sec^2\theta$, and $\beta'(X)=-1/h$.

\begin{table}[tbhp] 
\caption{Absorption coefficients $\mathscr{A}(\sigma,\theta)$ for the exponential profile.}
\texttt{
\begin{tabular}{@{\extracolsep{-5.5pt}}rrrrrrrrrrr@{\hspace{0pt}}}
\hline
 \multicolumn{1}{c}{$\sigma$} & \multicolumn{1}{c}{$\theta=0^\circ$}  & \multicolumn{1}{c}{$10^\circ$}& \multicolumn{1}{c}{$20^\circ$}& \multicolumn{1}{c}{$30^\circ$}& \multicolumn{1}{c}{$40^\circ$}& \multicolumn{1}{c}{$50^\circ$}& \multicolumn{1}{c}{$60^\circ$}& \multicolumn{1}{c}{$70^\circ$}& \multicolumn{1}{c}{$80^\circ$}& \multicolumn{1}{c}{$90^\circ$}\\ 
\hline 
0.0000 & 0.0000 & 0.0000 & 0.0000 & 0.0000 & 0.0000 & 0.0000 & 0.0000 &
   0.0000 & 0.0000 & 0.0000 \\
 0.0200 & 0.0510 & 0.0329 & 0.0100 & 0.0029 & 0.0010 & 0.0004 & 0.0001 &
   0.0000 & 0.0000 & 0.0000 \\
 0.0500 & 0.1209 & 0.1028 & 0.0606 & 0.0273 & 0.0114 & 0.0048 & 0.0020 &
   0.0007 & 0.0002 & 0.0000 \\
 0.1000 & 0.2207 & 0.2048 & 0.1602 & 0.1033 & 0.0572 & 0.0287 & 0.0131 &
   0.0050 & 0.0011 & 0.0000 \\
 0.2000 & 0.3653 & 0.3530 & 0.3163 & 0.2576 & 0.1868 & 0.1192 & 0.0654 &
   0.0283 & 0.0069 & 0.0000 \\
 0.3000 & 0.4495 & 0.4395 & 0.4089 & 0.3574 & 0.2873 & 0.2063 & 0.1266 &
   0.0598 & 0.0153 & 0.0000 \\
 0.4000 & 0.4877 & 0.4788 & 0.4516 & 0.4050 & 0.3387 & 0.2560 & 0.1654 &
   0.0809 & 0.0203 & 0.0000 \\
 0.4644 & 0.4937 & 0.4852 & 0.4591 & 0.4142 & 0.3497 & 0.2675 & 0.1745 &
   0.0852 & 0.0205 & 0.0000 \\
 0.5000 & 0.4921 & 0.4837 & 0.4579 & 0.4137 & 0.3499 & 0.2681 & 0.1749 &
   0.0846 & 0.0197 & 0.0000 \\
 0.6000 & 0.4729 & 0.4646 & 0.4393 & 0.3959 & 0.3336 & 0.2536 & 0.1623 &
   0.0749 & 0.0153 & 0.0000 \\
 0.7000 & 0.4386 & 0.4303 & 0.4050 & 0.3620 & 0.3010 & 0.2238 & 0.1378 &
   0.0589 & 0.0100 & 0.0000 \\
 0.8000 & 0.3955 & 0.3871 & 0.3620 & 0.3197 & 0.2606 & 0.1878 & 0.1096 &
   0.0425 & 0.0058 & 0.0000 \\
 0.9000 & 0.3485 & 0.3403 & 0.3156 & 0.2745 & 0.2184 & 0.1513 & 0.0828 &
   0.0286 & 0.0030 & 0.0000 \\
 1.0000 & 0.3011 & 0.2931 & 0.2693 & 0.2303 & 0.1781 & 0.1180 & 0.0600 &
   0.0181 & 0.0014 & 0.0000 \\
 1.1000 & 0.2558 & 0.2482 & 0.2257 & 0.1893 & 0.1419 & 0.0895 & 0.0420 &
   0.0110 & 0.0006 & 0.0000 \\
 1.2000 & 0.2140 & 0.2069 & 0.1860 & 0.1529 & 0.1108 & 0.0662 & 0.0284 &
   0.0064 & 0.0003 & 0.0000 \\
 1.3000 & 0.1767 & 0.1702 & 0.1511 & 0.1215 & 0.0849 & 0.0479 & 0.0187 &
   0.0036 & 0.0001 & 0.0000 \\
 1.4000 & 0.1440 & 0.1382 & 0.1212 & 0.0951 & 0.0640 & 0.0340 & 0.0121 &
   0.0019 & 0.0000 & 0.0000 \\
 1.6000 & 0.0925 & 0.0880 & 0.0751 & 0.0560 & 0.0348 & 0.0162 & 0.0047 &
   0.0005 & 0.0000 & 0.0000 \\
 1.8000 & 0.0571 & 0.0538 & 0.0445 & 0.0315 & 0.0179 & 0.0073 & 0.0017 &
   0.0001 & 0.0000 & 0.0000 \\
 2.0000 & 0.0339 & 0.0316 & 0.0254 & 0.0170 & 0.0088 & 0.0031 & 0.0005 &
   0.0000 & 0.0000 & 0.0000 \\
 2.2000 & 0.0195 & 0.0180 & 0.0140 & 0.0088 & 0.0041 & 0.0012 & 0.0002 &
   0.0000 & 0.0000 & 0.0000 \\
 2.4000 & 0.0109 & 0.0099 & 0.0074 & 0.0044 & 0.0018 & 0.0005 & 0.0000 &
   0.0000 & 0.0000 & 0.0000 \\
 2.6000 & 0.0059 & 0.0053 & 0.0038 & 0.0021 & 0.0008 & 0.0002 & 0.0000 &
   0.0000 & 0.0000 & 0.0000 \\
 2.8000 & 0.0031 & 0.0028 & 0.0019 & 0.0010 & 0.0003 & 0.0001 & 0.0000 &
   0.0000 & 0.0000 & 0.0000 \\
 3.0000 & 0.0016 & 0.0014 & 0.0009 & 0.0004 & 0.0001 & 0.0000 & 0.0000 &
   0.0000 & 0.0000 & 0.0000
\end{tabular} } \label{Asigmatheta}
\end{table}

\begin{table}[tbhp] 
\caption{Phase retardations $\Delta\phi(\sigma,\theta)$ (degrees) for the exponential profile. The absence of several entries at $\sigma=20$ is due to numerical difficulties.}
\texttt{
\begin{tabular}{@{\extracolsep{-3pt}}rrrrrrrrrr@{\hspace{0pt}}}
\hline
 \multicolumn{1}{c}{$\sigma$} & \multicolumn{1}{c}{$\theta=0^\circ$}  & \multicolumn{1}{c}{$10^\circ$}& \multicolumn{1}{c}{$20^\circ$}& \multicolumn{1}{c}{$30^\circ$}& \multicolumn{1}{c}{$40^\circ$}& \multicolumn{1}{c}{$50^\circ$}& \multicolumn{1}{c}{$60^\circ$}& \multicolumn{1}{c}{$70^\circ$}& \multicolumn{1}{c}{$80^\circ$}\\ 
\hline 
0.0000 & 0.00 & 0.00 & 0.00 & 0.00 & 0.00 & 0.00 & 0.00 &
   0.00 & 0.00 \\
 0.0200 & 2.65 & 2.58 & 2.20 & 1.79 & 1.49 & 1.29 & 1.16 &
   1.07 & 1.03 \\
 0.0500 & 6.71 & 6.70 & 6.44 & 5.88 & 5.29 & 4.78 & 4.41 &
   4.16 & 4.02 \\
 0.1000 & 13.69 & 13.74 & 13.71 & 13.37 & 12.82 & 12.21 &
   11.69 & 11.30 & 11.08 \\
 0.2000 & 28.36 & 28.50 & 28.80 & 29.05 & 29.12 & 29.09 &
   29.04 & 29.07 & 29.20 \\
 0.3000 & 43.53 & 43.74 & 44.30 & 45.03 & 45.78 & 46.54 &
   47.34 & 48.22 & 49.12 \\
 0.4000 & 58.45 & 58.72 & 59.47 & 60.58 & 61.90 & 63.36 &
   64.97 & 66.70 & 68.38 \\
 0.4644 & 67.57 & 67.87 & 68.72 & 70.01 & 71.63 & 73.47 &
   75.51 & 77.70 & 79.78 \\
 0.5000 & 72.36 & 72.68 & 73.58 & 74.96 & 76.71 & 78.73 &
   80.98 & 83.38 & 85.63 \\
 0.6000 & 84.81 & 85.15 & 86.16 & 87.73 & 89.78 & 92.19 &
   94.90 & 97.75 & 100.33 \\
 0.7000 & 95.65 & 96.01 & 97.08 & 98.78 & 101.02 & 103.70 &
   106.70 & 109.80 & 112.46 \\
 0.8000 & 104.97 & 105.35 & 106.45 & 108.23 & 110.59 & 113.42
   & 116.57 & 119.74 & 122.32 \\
 0.9000 & 112.97 & 113.35 & 114.47 & 116.29 & 118.71 & 121.61
   & 124.79 & 127.91 & 130.30 \\
 1.0000 & 119.84 & 120.23 & 121.36 & 123.18 & 125.61 & 128.51
   & 131.65 & 134.62 & 136.77 \\
 1.1000 & 125.79 & 126.17 & 127.30 & 129.11 & 131.51 & 134.35
   & 137.37 & 140.14 & 142.04 \\
 1.2000 & 130.97 & 131.34 & 132.44 & 134.22 & 136.57 & 139.32
   & 142.18 & 144.70 & 146.38 \\
 1.3000 & 135.49 & 135.85 & 136.93 & 138.66 & 140.93 & 143.56
   & 146.23 & 148.51 & 149.98 \\
 1.4000 & 139.46 & 139.82 & 140.86 & 142.53 & 144.71 & 147.18
   & 149.65 & 151.71 & 153.00 \\
 1.6000 & 146.07 & 146.40 & 147.36 & 148.88 & 150.83 & 152.99
   & 155.07 & 156.72 & 157.75 \\
 1.8000 & 151.27 & 151.56 & 152.43 & 153.80 & 155.51 & 157.36
   & 159.08 & 160.43 & 161.28 \\
 2.0000 & 155.40 & 155.66 & 156.44 & 157.63 & 159.12 & 160.69
   & 162.13 & 163.26 & 163.97 \\
 2.2000 & 158.70 & 158.94 & 159.62 & 160.67 & 161.95 & 163.29
   & 164.51 & 165.47 & 166.08 \\
 2.4000 & 161.37 & 161.58 & 162.18 & 163.09 & 164.20 & 165.35
   & 166.40 & 167.23 & 167.77 \\
 2.6000 & 163.55 & 163.73 & 164.26 & 165.05 & 166.02 & 167.02
   & 167.94 & 168.67 & 169.14 \\
 2.8000 & 165.34 & 165.50 & 165.96 & 166.66 & 167.51 & 168.39
   & 169.21 & 169.86 & 170.28 \\
 3.0000 & 166.82 & 166.97 & 167.38 & 168.00 & 168.75 & 169.54
   & 170.27 & 170.85 & 171.23 \\[4pt]
20.0000 & & & & 179.31 & 179.35 & 179.39 & 179.43 & 179.47 & 179.49
\end{tabular} } \label{Deltaphi}
\end{table}

Equation (\ref{ERAT}) is solved numerically (see Appendix \ref{app:gen} for details) subject to the boundary condition $\chi\to0$ as $\zeta\to\infty$ to find $A_1$ and $B_1$, and hence $\mathscr{A}$ and $\Delta\phi$, both regarded as functions of the independent parameters $\sigma$ and $\theta$. These are plotted in Figure \ref{fig:Asig} and tabulated in Tables \ref{Asigmatheta} and \ref{Deltaphi}. 

The first panel in Figure \ref{fig:Asig} illustrates that $\mathscr{A}\to0$ as $\sigma\to0$ or $\infty$ at all angles $\theta$. The peak occurs at $\sigma=\sigma_0=0.4644$, $A_0=0.4937$ for the RAT index $\mathscr{A}(\sigma)=\mathscr{A}(\sigma,0)$ plotted here as a dashed curve. The peak at all other $\theta$ is situated very close to $\sigma_0$ but reduces monotonically in height with increasing angle.

But why this humped structure of $\mathscr{A}(\sigma)$? Recall that the distance $X$ between the reflection and resonance points depends only on $\theta$ (for a given profile $\Omega^2$), and not on $\epsilon$. As $\epsilon$ decreases towards zero ($\sigma\to\infty$), the evanescent solution in Region III decays ever more quickly (see Equation \ref{chigen}), so energy tunnelling across the gap vanishes and with it $\mathscr{A}$. This is also consistent with our earlier observation that the fast and Alfv\'en modes decouple as $k_y\to0$ in Equation (\ref{xiDEs}). On the other hand, as $\epsilon\to\infty$ ($\sigma\to0$), we again see a decay to zero. This is because the length scale of the solution increases without bound and the Alfv\'en singularity becomes progressively less important. This is seen explicitly in Equation (\ref{RAT}) as $\sigma\to0$.

The second panel of Figure \ref{fig:Asig} displays the phase retardation over the range $0\le\sigma\le3$ and $0^\circ\le\theta\le85^\circ$, showing that it increases monotonically from 0 at $\sigma=0$ ($\kperp=0$) to $180^\circ$ as $\sigma\to\infty$ ($\kperp \to\infty$; this is confirmed by solution at much larger $\sigma$) and depends only weakly on $\theta$ (with $\sigma$ held fixed).

\subsection{ERAT Equation in Rational Form}

Equation (\ref{ERAT}) may be re-expressed as an equation with rational coefficients by making the further change of variable $t=u(\zeta)$, resulting in
\begin{equation}
\left(\nu\, t-1\right)^2\chi'' +\left(\nu-\frac{1}{t}\right)\chi'-\left[(1-\sigma\nu)t+\sigma\right]\chi = 0
\end{equation}
which has regular singular points at $t=0$ and $t=1/\nu$ ($x=X$ and $x=+\infty$ respectively) and an irregular singular point at $t=\infty$. We have written $\cos^2\theta=1-\sigma\nu$ here. Frobenius expansions about each of the regular singular points may be derived.

\end{document}